\documentclass[letterpaper,aps, prl, reprint, superscriptaddress]{revtex4-1}
\usepackage[T1]{fontenc}
\usepackage{color}
\usepackage{xcolor}
\usepackage{pdfcolmk}
\usepackage{verbatim}
\usepackage{mathtools}
\usepackage{amsmath}
\usepackage{amssymb}
\usepackage{float}
\usepackage{array}
\usepackage{mathrsfs}
\usepackage{microtype}

\makeatletter

\providecolor{lyxadded}{rgb}{0,0,1}
\providecolor{lyxdeleted}{rgb}{1,0,0}

\DeclareRobustCommand{\lyxsout}[1]{\ifx\\#1\else\sout{#1}\fi}

\usepackage[colorlinks,citecolor=blue,linkcolor=red]{hyperref}

\begin{document}
\title{Non-interacting fractional topological Stark insulator}

\author{Yi-Hong Chen}
 \affiliation{International Center for Quantum Materials and School of Physics, Peking University, Beijing 100871, China}
 \affiliation{Hefei National Laboratory, Hefei 230088, China}

\author{Si-Yuan Chen}
 \affiliation{International Center for Quantum Materials and School of Physics, Peking University, Beijing 100871, China}
 \affiliation{Hefei National Laboratory, Hefei 230088, China}

\author{Xin-Chi Zhou}
 \affiliation{International Center for Quantum Materials and School of Physics, Peking University, Beijing 100871, China}
 \affiliation{Hefei National Laboratory, Hefei 230088, China}

 \author{Xiong-Jun Liu}
 \email{Corresponding author:xiongjunliu@pku.edu.cn}
 \affiliation{International Center for Quantum Materials and School of Physics, Peking University, Beijing 100871, China}
 \affiliation{Hefei National Laboratory, Hefei 230088, China}
 \affiliation{International Quantum Academy, Shenzhen 518048, China}

\begin{abstract}
Fractional topological phases, such as the fractional quantum Hall state, usually rely on strong interactions to generate ground state degeneracy with gap protection and fractionalized topological response. Here, we propose a fractional topological phase without interaction in $(1+1)$-dimension, which is driven by the Stark localization on top of topological flat bands, different from the conventional mechanism of the strongly correlated fractional topological phases. A linear potential gradient applied to the flat bands drives the Stark localization, under which the Stark localized states may hybridize and leads to a new gap in the real space, dubbed the real space energy gap (RSEG). Unlike the integer topological band insulator obtained in the weak linear potential regime without closing the original bulk gap, the fractional topological Stark insulating phase is resulted from the RSEG when the linear potential gradient exceeds a critical value. We develop a theoretical formalism to characterize the fractional topological Stark insulator, and further show that the many-body state under topological pumping returns to the initial state only after multiple $2\pi$ periods of evolution, giving the fractional charge pumping, similar to that in fractional quantum Hall state. Finally, we propose how to realize the fractional topological Stark insulator in real experiment.
\end{abstract}

\maketitle

\textcolor{blue}{\em Introduction.}--Fractional topological phases are exotic states of matter beyond Laudau paradigm. A most prominent example is the two-dimensional (2D) fractional quantum Hall (FQH) effect \cite{tsui1982two}. The FQH state is characterized not by local order parameter, but by fractional topological invariants defined globally for the many-body wavefunction. For instance, the Laughlin type $1/(2p+1)$ FQH state \cite{laughlin1983anomalous} is characterized by many-body Chern number defined over insertion of $(2p+1)$ flux quanta \cite{niu1985quantized}, which determines the fractionally quantized Hall conductance through Laughlin's gauge argument \cite{laughlin1981quantized}. This is due to its $(2p+1)$-fold ground state degeneracy on a torus \cite{haldane1985many,wen1990ground}, and these ground states evolve into each other under flux insertion, and return to themselves when $(2p+1)$ flux quanta are inserted. The electron interactions gap out the otherwise gapless fractionally filled Landau level, protecting the ground state degeneracy and fractional charge pumping 
under flux insertion. Besides, fractional excitations of FQH state exhibit anyon statistics \cite{laughlin1983anomalous,arovas1984fractional,halperin1984statistics,greiter2024fractional} associated with the topological order \cite{wen2017colloquium}.

While the exploration of fractional topological phases has been focused on the interacting quantum systems, a nontrivial question is that, whether such phases can be realized in gapped systems without interaction? At the first glance this seems impossible, since in the free fermion systems, any gapped many-body ground state must have no degeneracy due to the Pauli exclusion principle, which however is necessary for the conventional fractional topological phases. In addition to the FQH states, the fractional topological charge pumping is also obtained in the topological charge density wave phases~\cite{grusdt2014realization,zeng2016fractional,taddia2017topological,petrescu2017precursor,li2017finite,gonzalez2019intertwined}, in which the ground state degeneracy is due to the broken discrete lattice symmetry, and also the topological solitons~\cite{jurgensen2023quantized}. In all these states the fractionalization is driven by particle-particle interactions.

In this work, we propose a $(1+1)$D non-interacting fractional topological insulating phase emerging from Stark localization \cite{wannier1960wave,wannier1962dynamics,kruchinin2018colloquium} due to a linear potential applied to topological flat bands, termed fractional topological Stark insulator (TSI). 
As the linear potential steepens, the bulk energy gap changes from Bloch band gap to real-space interband hybridization gap, leading to the transition from integer topological band insulator to fractional TSI. The many-body wavefunction of the fractional TSI, enforced by Stark localization, evolves into orthogonal states under periodic Hamiltonian modulation and returns to the initial state after multiple $2\pi$ modulation, similar to FQH, 
for which fractional charge pumping is achieved.
We develop a generic theory to characterize the fractional TSI, and propose an experimental realization in optical lattice. 
Our work unveils a novel mechanism of fractionalization in topological phases. 

\begin{figure}[tbp]
	\includegraphics[width=\columnwidth]{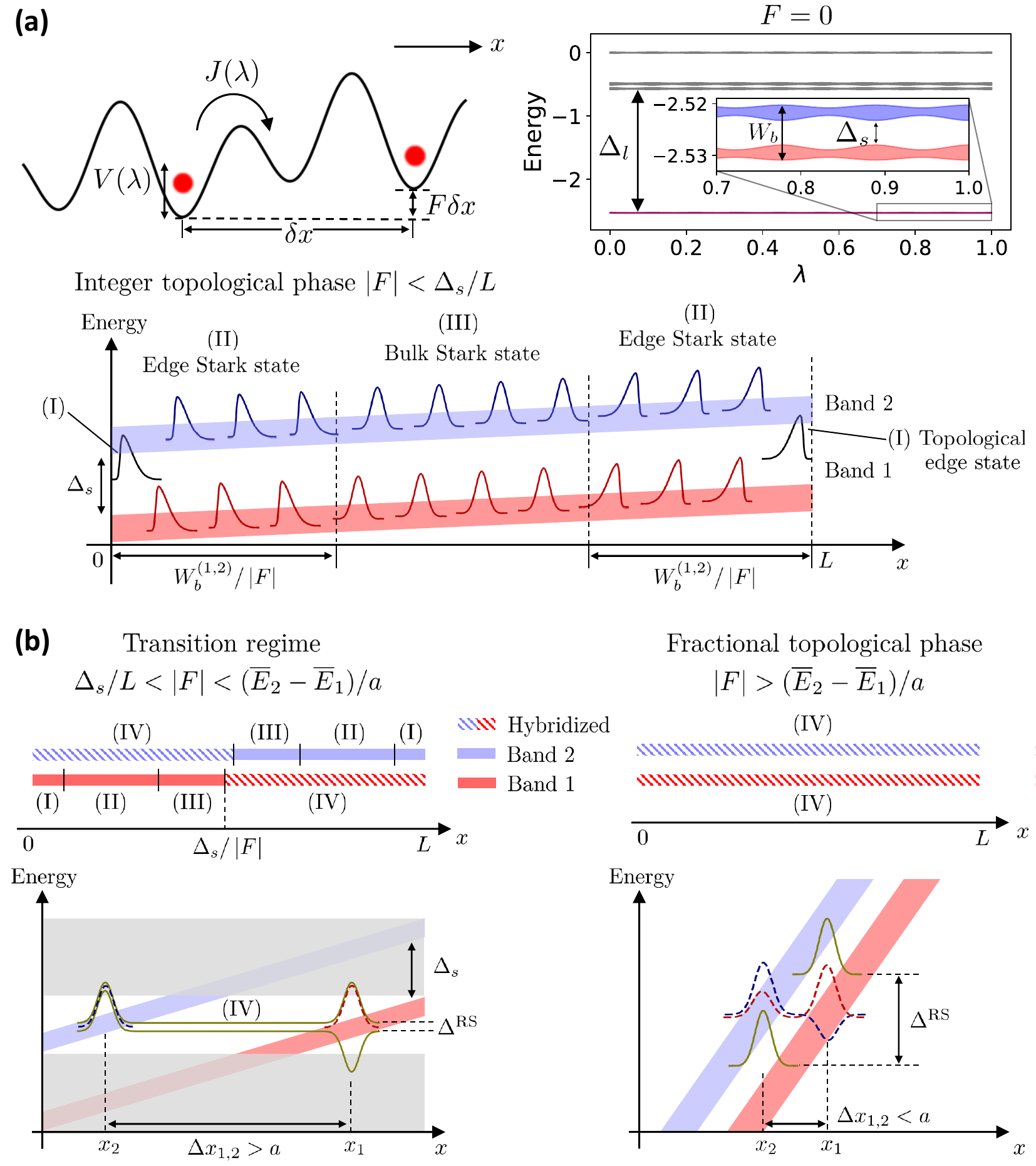}
	\caption{\label{Fig:real space energy gap}(a) Upper panel: (1+1)D topological insulator subject to a linear potential $F\hat{x}$. When $F=0$, the system hosts two lowest topological flat bands (filled with orange and blue) separated from higher bands (filled with gray) by a large band gap $\Delta_l$. Lower panel: The integer regime where Stark states from two bands never overlap in energy. Integer topology of each band is protected by gap $\Delta_s-\left|F\right|L$. (b) Left panel: transition regime where boundary integer region coexists with bulk fractional region. In the fractional region (\uppercase\expandafter{\romannumeral4}), Stark states from lower/upper band (red/blue) located at $x_{1,2}$ hybridize (green) and acquire a real space energy gap (RSEG) $\Delta^{\mathrm{RS}}$. The global topology is not well-defined. Right panel: fractional topological phase in strong linear potential regime with integer region disappearing. The states (dashed) from upper/lower band (blue/red) located in each unit cell are hybridized, giving multiband Stark states (solid). Fractional topology is protected by dominant RSEG $\Delta^{\mathrm{RS}}$. }
\end{figure}

\textcolor{blue}{\em General theory.}--We start with the generic theory for the (1+1)D noninteracting fractional TSI defined in one real-space dimension and one synthetic dimension characterized by a parameter $\lambda$ space, 
with the Hamiltonian
\begin{equation}
	H(\lambda)=H_0(\lambda)+F\sum_{m}\frac{x_m}{\mathcal{L}}c_{m}^{\dagger}c_{m}.
\end{equation}
Here $c_{m}^{\dagger}$ ($c_m$) is the fermion creation (annihilation) operator at the $m$-th lattice site (the real space position $x_m$), the unit cell size is set as $a=1$, $\mathcal{L}$ is a normalization factor, and $F$ characterizes the linear potential strength. The $\lambda$-parameterized Bloch Hamiltonians $H_0(\lambda)$ host $n_b$ topological flat bands [Fig. \ref{Fig:real space energy gap}(a)], which have the total Chern number $C$ defined in $(k_x,\lambda)$ space and are separated from higher bands by a large band gap $\Delta_l$ that is much larger than their total band width $W_b$. Without loss of generality, we consider the minimal case of $n_b=2$ and $C=1$, with a small gap $\Delta_s$ between the two flat bands. The applied linear potential offsets the band gap $\Delta_s$ by inclining the flat bands in real space and has important consequence upon increasing the strength $F$.

In the regime of a weak linear potential which does not close the original band gap $\Delta_s$, the system stays in the integer topological phase. As sketched in Fig. \ref{Fig:real space energy gap}(a), two Bloch bands turn into two Stark ladders comprising a series of Stark states under linear potential, with Stark states from different bands in the same unit cell separated by energy gap $\Delta_s$. For a system of length $L$, the two Stark ladders maintain their energy separation if $\left|F\right|<\Delta_s/L$. In this case, the many-body state formed by occupying all lower band Stark states is adiabatically connected to the integer topological phase without linear potential, and is protected by the gap $\Delta_s-\left|F\right|L$. In general, the system has three types of states: (\uppercase\expandafter{\romannumeral1}) topological edge states reminiscent of the band insulator, and (\uppercase\expandafter{\romannumeral3}) bulk Stark states which are uniform superpositions of all Bloch states. The energy of the bulk Stark state $|w_{j}^{(n)}\rangle$ of the $n$-th band centered at $j$-th unit cell is $\overline{E}_n+jFa$, with the position independent part being \cite{wannier1962dynamics,kruchinin2018colloquium}
\begin{equation}
    \overline{E}_n=\frac{1}{2\pi}\int_{0}^{2\pi} E_n(k_x)\mathrm{d}k_x + F\left(\langle w_{j}^{(n)}|\hat{x}|w_{j}^{(n)}\rangle-ja \right),
\end{equation}
where $E_n(k_x)$ is the dispersion of the $n$-th band. In addition, their localization lengths are given by $\xi^{(1,2)}\simeq W_b^{(1,2)}/\left|F\right|$ \cite{kruchinin2018colloquium}, where $W_b^{(1)}$ ($W_b^{(2)}$) is the width of the first (second) band. Within the distance $W_b^{(1,2)}/\left|F\right|$ from the boundary, the bulk Stark states cannot fully extend and turn into (\uppercase\expandafter{\romannumeral2}) edge Stark states. In region (\uppercase\expandafter{\romannumeral2}), particles are scattered off the boundary during Bloch oscillation, so they cannot traverse the whole Brillouin zone. As a result, the edge Stark states are only superpositions of a part of the Bloch states. Their energy is closer to band edge, and $\left|F\right|<\Delta_s/L$ guarantees that edge Stark states of different bands remain gapped.

\begin{figure}[tbp]
	\includegraphics[width=\columnwidth]{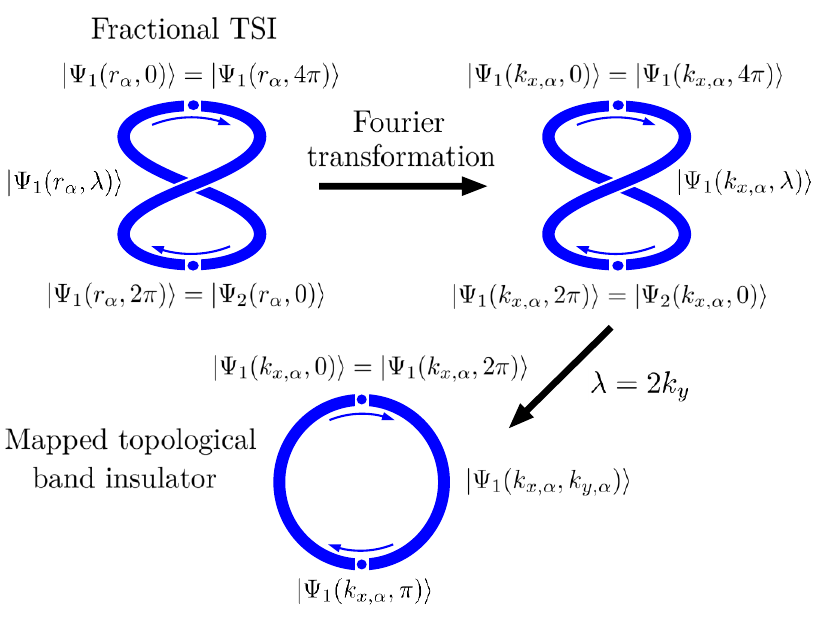}
	\caption{\label{Fig:fractional topology}Characterization of the fractional TSI. The fractional TSI state $\left|\Psi_1(x_{\alpha},\lambda=0)\right\rangle$ transitions into an orthogonal state $\left|\Psi_2(x_{\alpha},\lambda=0)\right\rangle$ for one cycle $\lambda=0\rightarrow\lambda=2\pi$ ($\alpha$ is particle label). It takes two cycles for initial state to return to itself, for which a many-body Berry phase can be defined. By first transforming the translationally invariant $\left|\Psi_1(x_{\alpha},\lambda)\right\rangle$ to momentum basis as $\left|\Psi_1(k_{x,\alpha},\lambda)\right\rangle$, and then mapping to synthetic dimension with doubled drive period $\lambda=2k_y$, the many-body Berry phase can be converted to the Chern number $C^{\prime}$ of a mapped artificial topological band insulator, which is calculated on the synthetic Brillouin zone spanned by $(k_x,k_y)$.
    }
\end{figure}

\begin{figure}[tbp]
	\includegraphics[width=\columnwidth]{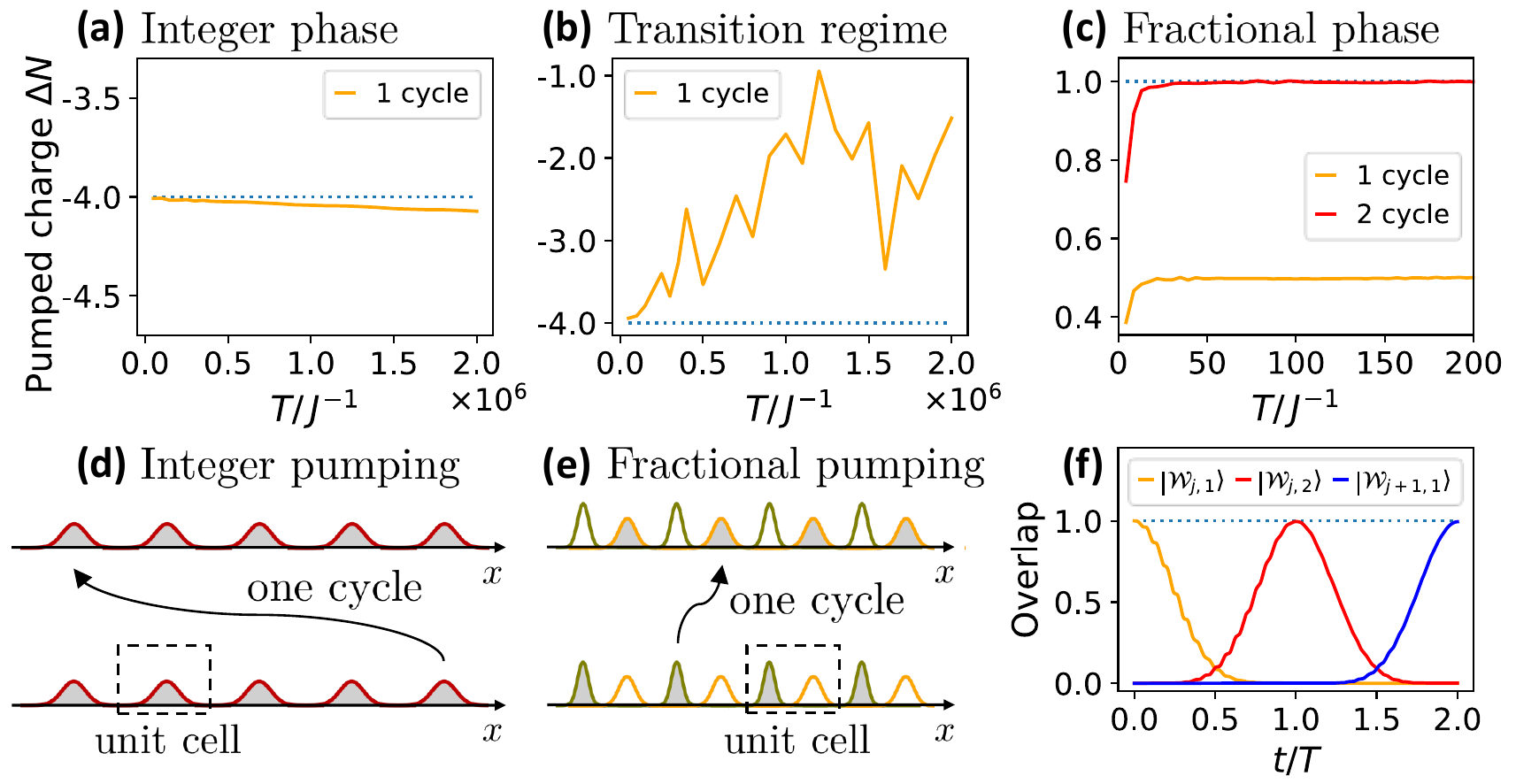}
	\caption{\label{fig:pumping}(a) In the integer topological phase, one-cycle pumped charge is quantized to $C_1=-4$. (b) In the transition regime, global topology is not well defined, and the charge pumping is not quantized. (c) In the fractional topological phase, two-cycle pumped charge is quantized to $C=1$, giving the $1/2$ fractionally quantized charge pumping for one cycle. (d-e) Sketch of wavefunction evolution during charge pumping, where shade indicates occupied Stark states. (d) Integer charge pumping of topological band insulator. All Stark states of the first band are occupied, and each of them moves by $C_1=-4$ unit cells over one cycle. (e) Fractional charge pumping of the Stark insulator, where color indicates different species of multiband Stark states. Only one species of multiband Stark states are occupied, and they evolve into the orthogonal ones after one cycle. (f) Overlap of the evolving state $U(t)\left|{\cal W}_{j,1}\right\rangle$ with static Stark states for $T=50J^{-1}$, indicating that $U(T)\left|{\cal W}_{j,1}\right\rangle=\left|{\cal W}_{j,2}(\lambda=0)\right\rangle$ and $U(2T)\left|{\cal W}_{j,1}\right\rangle=\left|{\cal W}_{j+1,1}(\lambda=0)\right\rangle$.}
\end{figure}

A novel fractional topological phase is obtained when the linear potential is tuned to a strong regime with $\left|F\right|>(\overline{E}_2-\overline{E}_1)/a$, in which regime two Stark ladders completely overlap in energy and hybridize [Fig. \ref{Fig:real space energy gap}(b), Right]. In this case, the degenerate bulk Stark states $|w_{j_1}^{(1)}\rangle$ $(|w_{j_2}^{(2)}\rangle)$ from lower (upper) band located at $x_{1}$ ($x_2$) are coupled by the linear potential, resulting in a dominant real space energy gap (RSEG) as $\Delta^{\mathrm{RS}}=2|\langle w_{j_1}^{(1)}|F\hat{x}|w_{j_2}^{(2)}\rangle|$ separating the hybridized states, which is essentially the linear potential difference between them. 
The distance satisfies $\Delta x_{1,2}=\left|x_1-x_2\right|=(\overline{E}_2-\overline{E}_1)/\left|F\right|\leq a$. Consequently, all bulk Stark states are hybridized with a degenerate partner in the same unit cell, so do the edge Stark states which have smaller energy gap $\Delta_s$, and integer regions (I,II,III) disappear. We term the hybridized states (\uppercase\expandafter{\romannumeral4}) {\em multiband Stark states} and denote them as $\left|{\cal W}_{j,\mu}\right\rangle$ with unit cell index $j$ and species index $\mu=1,2$. The multiband Stark states vary under the continuous parameter flow $\lambda=0\rightarrow \lambda=2\pi$ [Fig. \ref{fig:pumping}(e)] as
\begin{equation}\label{Eq:1/2 WS state evolution}
	\left|{\cal W}_{j,1}\right\rangle\rightarrow\left|{\cal W}_{j,2}\right\rangle,
    \left|{\cal W}_{j,2}\right\rangle\rightarrow\left|{\cal W}_{j+1,1}\right\rangle.
\end{equation}
Evolution (\ref{Eq:1/2 WS state evolution}) can be proved as follows. If all $\left|{\cal W}_{j,\mu=1,2}\right\rangle$ are occupied, the resultant many-body state yields the total $C=1$ charge pumping over one cycle. Confined to a unit cell, charge pumping implies the displacement of $\left|{\cal W}_{j,1}\right\rangle$ and $\left|{\cal W}_{j,2}\right\rangle$ adds up to one unit cell. However, since $\left|{\cal W}_{j,\mu}\right\rangle$ are gapped by RSEG, their energy ordering must be preserved, leaving (\ref{Eq:1/2 WS state evolution}) the only possible adiabatic evolution (see Supplementary material \ref{Supsec:multiband stark states}). With the multiband Stark states, we can obtain an exotic fractional TSI by fully occupying only one species of them, i.e.
\begin{equation}\label{Eq:many-body wavefunction}
    \left|\Psi_{\mu}(x_{\alpha})\right\rangle= \left|{\cal W}_{1,\mu},{\cal W}_{2,\mu},\dots,{\cal W}_{j,\mu},\dots\right\rangle,
\end{equation}
which is protected by the RSEG. 
According to Eq. (\ref{Eq:1/2 WS state evolution}), the many-body state $\left|\Psi_1(x_{\alpha})\right\rangle$ evolves into an orthogonal state $\left|\Psi_2(x_{\alpha})\right\rangle$ after one cycle ($\lambda=0\rightarrow \lambda=2\pi$), and returns to itself after two cycles. The associated many-body Berry phase leads to fractional topology as follows [Fig. \ref{Fig:fractional topology}]. Leveraging the translational invariance of $\left|\Psi_{\mu}(x_{\alpha})\right\rangle$, we define the many-body state in momentum space for an artificial band insulator through the artificial Bloch states $\left|{\cal V}_{k_x,\mu}(\lambda)\right\rangle=\sum_{j}\mathrm{e}^{\mathrm{i}k_x j}\left|{\cal W}_{j,\mu}(\lambda)\right\rangle/\mathcal{N}$ ($\mathcal{N}$ is normalization factor) at each $\lambda$, as given by
\begin{equation}
    \left|\Psi_{\mu}(k_{x,\alpha},\lambda)\right\rangle= \left|{\cal V}_{k_x=0,\mu},\dots,{\cal V}_{k_x,\mu}\dots,{\cal V}_{k_x=2\pi,\mu}\right\rangle.
\end{equation}
This state is periodic for $\lambda$ changing by $4\pi$, hence different from the many-body state in integer regime. By redefining $\lambda=2k_y$, the Berry phase of many-body state $\left|\Psi_{\mu}(k_{x,\alpha},k_{y,\alpha}=\lambda/2)\right\rangle$ gives the Chern number $C^{\prime}$ of the mapped artificial band insulator, as calculated in the synthetic Brillouin zone and equals the total Chern number $C$ of the two bands (see Supplementary Material \ref{SupSec:Chern number}),
\begin{equation}
	C^{\prime}=\frac{1}{2\pi}\int_{k_x=0}^{2\pi}\int_{k_y=0}^{2\pi}\tilde{\Omega}_{yx} \mathrm{d}k_x\mathrm{d}k_y=C,
\end{equation}
where the Berry curvature is given by $\tilde{\Omega}_{yx}=-2\mathrm{Im}\left\langle \partial_{k_y}{\cal V}_{k_x,k_y}|\partial_{k_x}{\cal V}_{k_x,k_y}\right \rangle$. Thus the Stark insulator exhibits $C/2=1/2$ fractional topology, since the Chern number is only defined for two cycles together, similar to that defined for FQH state \cite{niu1985quantized}. However, the two many-body states evolving into each other, as the key ingredient of fractional topology, originates in the interplay between Stark localization and relative fractional filling of the multiband Hilbert space. These two states are not degenerate, and they attain gap protection from Stark localization. This is significantly different from FQH states, where the fractionalization and gap protection are induced by strong interactions.

The transition regime between integer and fractional phases is obtained for $\Delta_s/L<\left|F\right|<(\overline{E}_2-\overline{E}_1)/a$ [see Fig. \ref{Fig:real space energy gap}(b), Left]. In the region (\uppercase\expandafter{\romannumeral4}), the coupled degenerate bulk Stark states are spatially separated apart by $\left|x_1-x_2\right|=(\overline{E}_2-\overline{E}_1)/\left|F\right|>a$, and RSEG $\Delta^{\mathrm{RS}}$ is small. Meanwhile, within the distance $\Delta_s/\left|F\right|$ from the boundary, Stark states lack degenerate partners for hybridization, thereby preserving integer regions (\uppercase\expandafter{\romannumeral1}-\uppercase\expandafter{\romannumeral3}) close to boundary. The coexisting regions without overall bulk gap preclude well-defined global topology. Further, in the thermodynamic limit $L\rightarrow\infty$, the transition occurs at $\left|F\right|\rightarrow0$. The integer topological phase only exists at $\left|F\right|=0$, while fractional TSI is obtained for finite $\left|F\right|>0$. 
Note that the transition is associated with the closing of RSEG $\Delta^{\mathrm{RS}}$, with the scaling
\begin{equation}
    \begin{split}
        &\xi^{(1,2)}\simeq \frac{W_b^{(1,2)}}{\left|F\right|},\,\Delta x_{1,2}=\frac{\overline{E}_2-\overline{E}_1}{\left|F\right|},\\
        \Delta^{\mathrm{RS}}&=2\left|\langle w_{j_1}^{(1)}|F\hat{x}|w_{j_2}^{(2)}\rangle\right| \sim \left|F\right|\exp\left(-\frac{\Delta x_{1,2}}{\xi^{(1,2)}}\right)
    \end{split}
\end{equation}
in the $\left|F\right|\rightarrow0$ limit, where the exponential factor originates from the overlap between localized Stark states (see Supplementary Material \ref{SupSec:RSEG scaling}). We observe that $\Delta^{\mathrm{RS}}$ scales linearly as $\left|F\right|$, by noticing $\xi^{(1,2)}$ and $\Delta x_{1,2}$ both scale as $\left|F\right|^{-1}$, rendering a trivial exponential factor. The fractional TSI obtained for $\left|F\right|>0$ is attributed to the divergence of the linear potential energy $\left|F\right|L\rightarrow\infty$ for any non-zero $\left|F\right|$ in the thermodynamic limit.

These results can be readily extended to $n_b$ topological flat bands with total Chern number $C$, far separated from the remaining bands. In the fractional phase, multiband Stark states $\left|{\cal W}_{j,\mu}\right\rangle$ ($\mu=1,2,\dots,n_b$) are given in the strong linear potential regime and protected by RSEG $\Delta^{\mathrm{RS}}$. The many-body state of the fractional TSI is constructed in the same way as in Eq. (\ref{Eq:many-body wavefunction}), evolves into orthogonal states by varying $\lambda$, and returns to itself after $2\pi n_b$ modulation, rendering the $C/n_b$ fractional number for the phase (see Supplementary material \ref{Supsec:multiband stark states}).

\textcolor{blue}{\em Fractional topological pumping.}--The topology of the fractional TSI can be probed through $C/n_b$ fractionally quantized charge pumping, in which the transferred charge is topologically quantized to $C$ only for $n_b$ drive cycles together (see Supplementary material \ref{SupSec:fractional pump}). We exemplify with a $1/2$ fractional TSI whose Hamiltonian includes a $2/9$ commensurate generalized Aubry-Andr\'{e} (GAA) model with modulated hopping, $H_0(\lambda)=\sum_{m}\left[ J+\delta J\cos\left(\lambda-\frac{4\pi}{9}m\right) \right]c_{m+1}^{\dagger}c_{m}+\mathrm{h.c.}$, with $J=-1,\,\delta J=-0.98$, and $\lambda(t)$ tuned by time $t$. The lowest two bands carry Chern numbers $C_1=-4,\,C_2=5$, adding up to a total Chern number $C=1$, separated from the third band by a gap $\Delta_l=1.94$ much greater than their band width $W_b=0.011$. According to the general theory, the system hosts a $1/2$ fractional TSI under strong linear potential. We probe the topology in different regimes of the integer-fractional transition in Fig. \ref{fig:pumping}(a-c) with charge pumping. The many-body state is initialized by occupying the first band in the integer region and one species of multiband Stark states in the fractional region. Fig. \ref{fig:pumping}(a) shows integer quantization plateau for $C_1=-4$. In this case, all Stark states of the first band are occupied and move by $C_1=-4$ unit cells over one cycle [Fig. \ref{fig:pumping}(d)], leading to the $\Delta N=-4$ integer charge pumping for the many-body state. In the fractional region Fig. \ref{fig:pumping}(c), charge pumping is quantized to $C=1$ for $n_b=2$ cycles but is fractional $\Delta N=1/2$ for one cycle. After one cycle the many-body state evolves into an orthogonal state characterized by a different species of multiband Stark states being occupied, as shown in Fig. \ref{fig:pumping}(e), and returns to itself after two cycles. We also confirm the adiabatic change of Eq. (\ref{Eq:1/2 WS state evolution}) in Fig. \ref{fig:pumping}(f) by computing the overlap of the evolved state $U(t)\left|{\cal W}_{j,1}\right\rangle$ with static Stark states $\left|{\cal W}_{j,1}(\lambda=0)\right\rangle$, $\left|{\cal W}_{j,2}(\lambda=0)\right\rangle$ and $\left|{\cal W}_{j+1,1}(\lambda=0)\right\rangle$. However, in the transition regime Fig. \ref{fig:pumping}(b) where integer and fractional regions coexist, charge pumping depends on details of the system and is not topologically quantized. In addition, we probe the fractional topology of a $1/3$ fractional TSI through charge pumping in Supplementary Material \ref{SupSec:1/3 pumping}.

\begin{figure}[tbp]
	\includegraphics[width=\columnwidth]{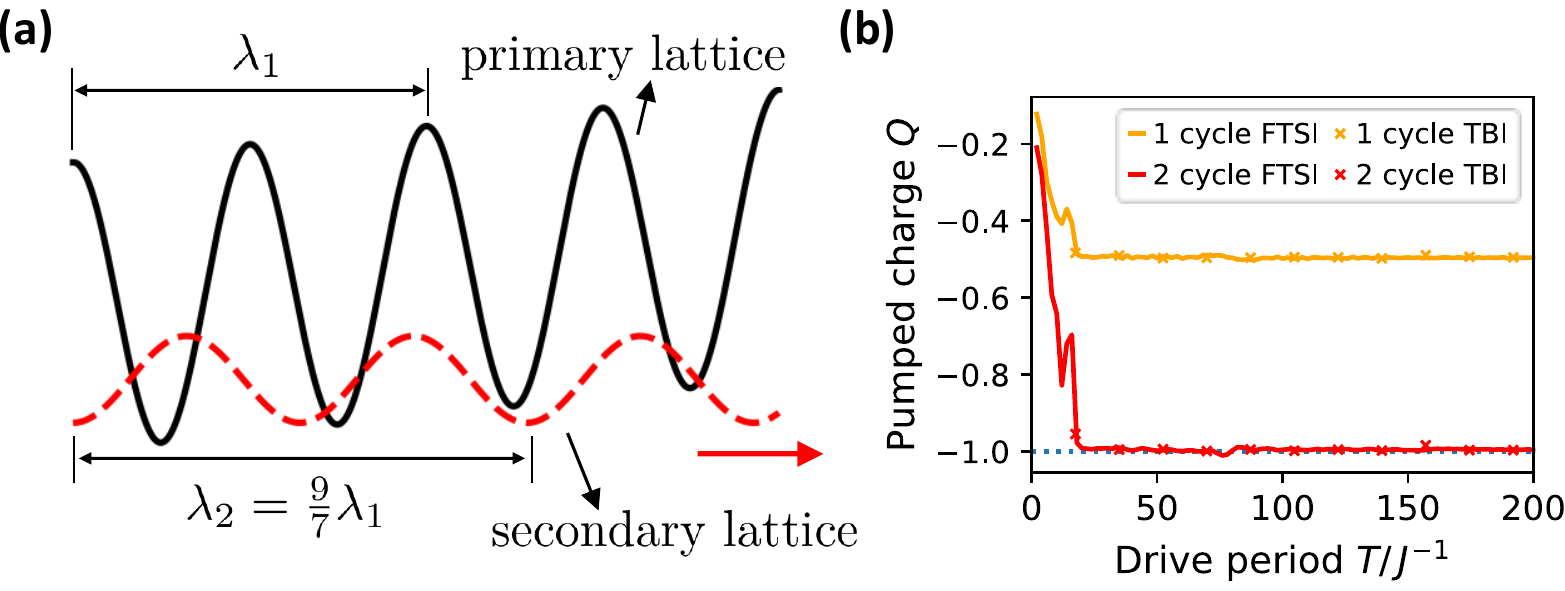}
	\caption{\label{Fig:experiment}Experimental setup and numerics. (a) The bichromatic lattice with linear potential. The deep primary lattice gives the tight-binding lattice, while the shallow secondary lattice provides on-site energy modulation and moves to the right as we increase $\lambda$ in the driving. (b) Charge pumping for the experimental setup. Two-cycle quantization plateau is observed at driving period $T>20J^{-1}$, both for fractional TSI (FTSI, solid line) and topological band insulator (TBI, x-marker) with commensurate driving period $T=kT_{\mathrm{Bloch}}$.}
\end{figure}

\textcolor{blue}{\em Experimental proposal.}--Finally we propose an experimental scheme to realize the fractional TSI with ultracold atoms in optical lattice. The topological flat bands can be obtained by imposing a bichromatic lattice [Fig. \ref{Fig:experiment}(a)] $H^{\mathrm{OL}}_0(\lambda)=\hat{p}_x^2/2m_a + V_1 \sin^2 \left( k_1 x \right) + V_2 \sin^2\left( k_2 x -\lambda \right)$, where $m_a$ is the atom mass, $k_i=2\pi/\lambda_i\,(i=1,2)$ are lattice wavenumbers giving recoil energy $E_{R,i}=\hbar^2 k_i^2 / 2 m_a$, and $\lambda$ is the driving parameter tuned by the phase of the light. We set $V_1\gg V_2$, such that laser of wavelength $\lambda_1$ creates a primary lattice with period $d=\lambda_1/2$ in the tight-binding limit, perturbed by the secondary lattice generated by laser of wavelength $\lambda_2$. The resultant tight-binding Hamiltonian is an Aubry-Andr\'{e} (AA) model with a variety of topological flat bands \cite{roati2008anderson,modugno2009exponential,schreiber2015observation} $H_{0}(\lambda)=\sum_{m}(Jc_{m+1}^{\dagger}c_{m}+\mathrm{h.c.})+\sum_{m}V\cos\left( \lambda-2\pi\beta m\right)c_{m}^{\dagger}c_{m}$,
in which $\beta=\lambda_1/\lambda_2$ is set to a rational number (details in Supplementary material \ref{SupSec:experimental parameter}). The linear potential can be generated and tuned by magnetic field gradient \cite{simon2011quantum,aidelsburger2013realization,miyake2013realizing}. Eventually, the charge pumping can be probed by measuring center-of-mass shift of the atom cloud \cite{lohse2016thouless,nakajima2016topological,nakajima2021competition}.

We take $^{40}\mathrm{K}$ atoms as an example. The wavelengths are taken as $\lambda_1=827.5\,\mathrm{nm}$ and $\lambda_2=1064\,\mathrm{nm}$, which yields a commensurate $\beta=7/9$, the recoil energy $E_{R,1}=7.3\,\mathrm{kHz}$ and unit cell length $a=9d=7.4\,\mu\mathrm{m}$. For lattice depth $V_1/E_{R,1}=10$ and $V_2/E_{R,2}=0.13$, it is estimated that $J\simeq 0.02E_{R,1}=145\,\mathrm{Hz}$, $V\simeq 2J$, and the system lifetime $\tau\simeq 10.5\,\mathrm{s}\simeq 1500J^{-1}$ (see details in Supplementary material \ref{SupSec:experimental parameter}). In this setup, the lowest two bands are separated from the third band by gap $\Delta_l=1.6J$, significantly greater than their own band width $W_b=0.08J$, and they carry total Chern number $C=-1$. We set linear potential to $Fa=0.72J$ satisfying $\Delta_l>Fa\gg W_b>\overline{E}_2-\overline{E}_1$, such that the system enters fractional phase characterized by $-1/2$ fractional number. The fractional TSI exhibits clear fractional quantization of charge pumping for drive period $T>20J^{-1}$ in Fig. \ref{Fig:experiment}(b), which can be observed within system lifetime. We note that the fractional charge pumping can probed even without precisely preparing the state in the many-body on as Eq.~\eqref{Eq:many-body wavefunction}. In Supplementary material \ref{SupSec:initial state}, we show that with a generic initialization which is a superposition of two species of multi-band Stark states, the fractional charge pumping corresponding to the fractional TSI phase is achieved when the driven period of pumping is integer times of Bloch oscillation period $T_{\mathrm{Bloch}}$.

\textcolor{blue}{\em Conclusion.}--We have proposed a non-interacting fractional topological Stark insulator (TSI) in the $(1+1)$D space, without topological order or spontaneous symmetry breaking. Applying linear potential to a set of topological flat bands opens up a new real space gap which dominates in the strong Stark localization regime, and drives the system to turn from integer topological band insulator to fractional TSI.
The fractional TSI can be characterized by the fractional many-body Berry phase, and detected by fractional charge pumping in experiment. Our findings unveil an exotic mechanism of fractional topological phase in non-interacting system: the Stark localization plays the role of interactions that induce the bulk gap and multiplet of states generating fractionalization, and can be extended to realize fractional symmetry-protected \cite{stern2016fractional,bernevig2006quantum,levin2009fractional,maciejko2010fractional,swingle2011correlated} and higher-dimensional \cite{zhang2001four,lohse2018exploring,zilberberg2018photonic,bouhiron2024realization} topological phases without interactions. This work opens a broad avenue to explore exotic fractional topological phases, being not only important in theory, but also feasible in experiment. There are also intriguing open questions deserving to be answered and explored in the future, including the statistics of the excitations in the fractional TSI phase.

\textcolor{blue}{\em Acknowledgement.} We thank Hongyu Wang for helpful discussion. This work was supported by National Key Research and Development Program of China (2021YFA1400900), the National Natural Science Foundation of China (Grants No. 12425401 and No. 12261160368), the Innovation Program for Quantum Science and Technology (Grant No. 2021ZD0302000), and the Shanghai Municipal Science and Technology Major Project (Grant No.~2019SHZDZX01).

\bibliography{ref}

\renewcommand{\thesection}{S-\arabic{section}}
\setcounter{section}{0}  
\renewcommand{\theequation}{S\arabic{equation}}
\setcounter{equation}{0}  
\renewcommand{\thefigure}{S\arabic{figure}}
\setcounter{figure}{0}  
\renewcommand{\thetable}{S\Roman{table}}
\setcounter{table}{0}  
\onecolumngrid \flushbottom 

\newpage
\begin{center}
	\large \textbf{\large Supplementary Material}
\end{center}

\section{The generic multiband Stark states}\label{Supsec:multiband stark states}

In this section, we detail the properties of the multiband Stark states $\left|{\cal W}_{j,\mu}\right\rangle$ for general case with $n_b$ bands with total Chern number $C$, especially their adiabatic evolution over one cycle
\begin{equation}\label{Eq:general WS state evolution}
    \left|{\cal W}_{j,\mu}\right\rangle \rightarrow \left|{\cal W}_{j+\lfloor (\mu+C-1)/n_b\rfloor,1+(\mu+C-1)\mathrm{mod} n_b}\right\rangle,\,\mu=1,2,\dots,n_b,
\end{equation}
which generalizes Eq. (\ref{Eq:1/2 WS state evolution}), with $\lfloor\,\,\rfloor$ being the round-down function. As explained in the main text, Eq. (\ref{Eq:general WS state evolution}) originates in the interplay between band topology and RSEG induced by Stark localization. Here we provide a more detailed and rigorous proof in the same spirit. First, we consider the infinite linear potential limit $\left|F\right|\rightarrow\infty$, where multiband Stark states turn into {\em maximally localized Wannier function} (MLWF), for which Eq. (\ref{Eq:general WS state evolution}) is obtained through analytic derivation. Second, the multiband Stark states at $\left|F\right|\rightarrow\infty$ and finite $\left|F\right|$ can be adiabatically connected, so adiabatic evolution (\ref{Eq:general WS state evolution}) is preserved.

In order to approach the $\left|F\right|\rightarrow\infty$ limit, we consider Hamiltonian
\begin{equation}\label{Eq:interpolation Hamiltonian}
		\tilde{H}(\lambda,\alpha)=P(F\hat{x}+\alpha H_0(\lambda))P=\alpha P(F_{\mathrm{eff}}\hat{x}+H_0(\lambda))P,\,F_{\mathrm{eff}}=F/\alpha,
\end{equation}
where $P$ is projection onto the $n_b$ bands, and parameter $\alpha$ controls the effective linear potential gradient $F_{\mathrm{eff}}$. In the limit $\alpha=0,\,\left|F_{\mathrm{eff}}\right|\rightarrow\infty$, the Hamiltonian reads $\tilde{H}(\lambda,0)=PF\hat{x}P=FH_{\mathrm{MLWF}}$, where $H_{\mathrm{MLWF}}=P\hat{x}P$ is known to give MLWF as eigenstates \cite{marzari1997maximally,fidkowski2011model,alexandradinata2014wilson}, denoted as $\left|{\cal W}^{\mathrm{MLWF}}_{j,\mu}\right\rangle=\left|{\cal W}_{j,\mu}(\alpha=0)\right\rangle$. MLWFs can also be obtained from Brillouin zone Wilson loop \cite{fidkowski2011model,alexandradinata2014wilson}, thus bridging Stark localization and band topology. The Wilson loop is defined as
\begin{equation}
	W_{k_x^{(0)}+2\pi\leftarrow k_x^{(0)}}(\lambda)=\mathcal{P}\exp \left[\mathrm{i}\int_{k_x^{(0)}}^{k_x^{(0)}+2\pi}\mathbf{A}_x\mathrm{d}k_x\right],
\end{equation}
where $\mathcal{P}$ refers to path ordering, and $\mathbf{A}_x$ is the non-Abelian (multi-band) Berry connection
\begin{equation}
	\mathbf{A}_x^{mn}=\left\langle u^m_{k_x}(\lambda) \right| \mathrm{i}\partial_{k_x} \left| u^n_{k_x}(\lambda) \right \rangle.
\end{equation}
Here $| u^{m,n}_{k_x}(\lambda) \rangle$ are Bloch states with band label $m,n=1,\dots,n_b$. Eigenstates of Wilson loop are $n_b$ Bloch states at $(k_x^{(0)},\lambda)$, and we denote them as $\left| \tilde{u}^{\mu}_{k_x^{(0)}}(\lambda)  \right\rangle,\,\mu=1,\dots,n_b$,
\begin{equation}
	W_{k_x^{(0)}+2\pi\leftarrow k_x^{(0)}}(\lambda) \left| \tilde{u}^{\mu}_{k_x^{(0)}}(\lambda)  \right\rangle = \mathrm{e}^{2\pi\mathrm{i}\phi_{\mu}(\lambda)}\left| \tilde{u}^{\mu}_{k_x^{(0)}}(\lambda)  \right\rangle.
\end{equation}
We can construct other Bloch states from parallel transport and enforce periodic gauge,
\begin{equation}\label{Eq:enforce periodic}
	| \tilde{u}^{\mu}_{k_x}(\lambda)  \rangle=\mathrm{e}^{-\mathrm{i}(k_x-k_x^{(0)})\phi_{\mu}}\mathcal{P}\exp \left[\mathrm{i}\int_{k_x^{(0)}}^{k_x}\mathbf{A}_x\mathrm{d}k_x\right]\left| \tilde{u}^{\mu}_{k_x^{(0)}}(\lambda)  \right\rangle.
\end{equation}
These Bloch states satisfy differential equation $| \tilde{u}^{\mu}_{k_x}(\lambda)  \rangle=(\phi_{\mu}-\mathbf{A}_x)| \tilde{u}^{\mu}_{k_x}(\lambda)  \rangle$. After Fourier transformation, we obtain MLWFs $\left|{\cal W}^{\mathrm{MLWF}}_{j,\mu}\right\rangle=\int\frac{\mathrm{d}k_x}{2\pi}\mathrm{e}^{-\mathrm{i}k_x j}| \tilde{u}^{\mu}_{k_x}(\lambda)  \rangle$, which can be verified to satisfy
\begin{equation}
	P\hat{x}P\left|{\cal W}^{\mathrm{MLWF}}_{j,\mu}\right\rangle=(\mathrm{i}\partial_{k_x}+\mathbf{A}_x)\left|{\cal W}^{\mathrm{MLWF}}_{j,\mu}\right\rangle=(\phi_{\mu}+j)\left|{\cal W}^{\mathrm{MLWF}}_{j,\mu}\right\rangle,
\end{equation}
so their $P\hat{x}P$ eigenvalues or MLWF centers are $\{\phi_{\mu}\}$.

Second, the $P\hat{x}P$ eigenvalues $\{\phi_{\mu}\}$ flow according to total Chern number $C$ under periodic drive $\lambda=0\rightarrow \lambda=2\pi$. Note that
\begin{equation}
	\exp\left[\sum_{\mu}2\pi\mathrm{i}\phi_{\mu}(\lambda)\right]=\det W_{k_x^{(0)}+2\pi\leftarrow k_x^{(0)}}(\lambda)=\exp\left[\mathrm{i}\int\mathrm{tr}\mathbf{A}_x\mathrm{d}k_x\right]=\exp\left[\mathrm{i}\sum_{n=1}^{n_b}\int \mathbf{A}_x^{nn}\mathrm{d}k_x\right].
\end{equation}
The last expression is exponential of sum of single-band Zak phases. We know that winding of each Zak phase gives band Chern number as $\lambda=0\rightarrow \lambda=2\pi$. To be explicit,
\begin{equation}
	\begin{split}
		2\pi\sum_{\mu}\left[ \phi_{\mu}(2\pi)-\phi_{\mu}(0)\right] &=\sum_{n=1}^{n_b}\int_{k_x=0}^{2\pi}\int_{\lambda=0}^{2\pi} \partial_{\lambda}\mathbf{A}_x^{nn}\mathrm{d}k_x \mathrm{d}\lambda\\
		&=\sum_{n=1}^{n_b}\int_{k_x=0}^{2\pi}\int_{\lambda=0}^{2\pi} \left[ \partial_{\lambda}\mathbf{A}_x^{nn}-\partial_{k_x}\mathbf{A}_{\lambda}^{nn}\right] \mathrm{d}k_x \mathrm{d}\lambda\\
		&=\sum_{n=1}^{n_b}\int_{k_x=0}^{2\pi}\int_{\lambda=0}^{2\pi} \Omega_{\lambda x}^{n} \mathrm{d}k_x \mathrm{d}\lambda\\
		&=2\pi \sum_{n=1}^{n_b} C_n=2\pi C,
	\end{split}
\end{equation}
with $C_n$ being Chern number of band $n$. Here we use the periodicity of $\mathbf{A}_{\lambda}^{nn}$ in $k_x$ direction, which we enforced by hand in Eq. (\ref{Eq:enforce periodic}), and single-band Berry curvature is defined as
\begin{equation}
	\Omega_{\lambda x}^{n}=\partial_{\lambda}\mathbf{A}_x^{nn}-\partial_{k_x}\mathbf{A}_{\lambda}^{nn}.
\end{equation}
So total change of MLWF center is determined by Chern number
\begin{equation}
	\sum_{\mu}\left[\phi_{\mu}(2\pi)-\phi_{\mu}(0)\right]=C.
\end{equation}

\begin{figure}[htbp]
	\includegraphics[width=0.8\columnwidth]{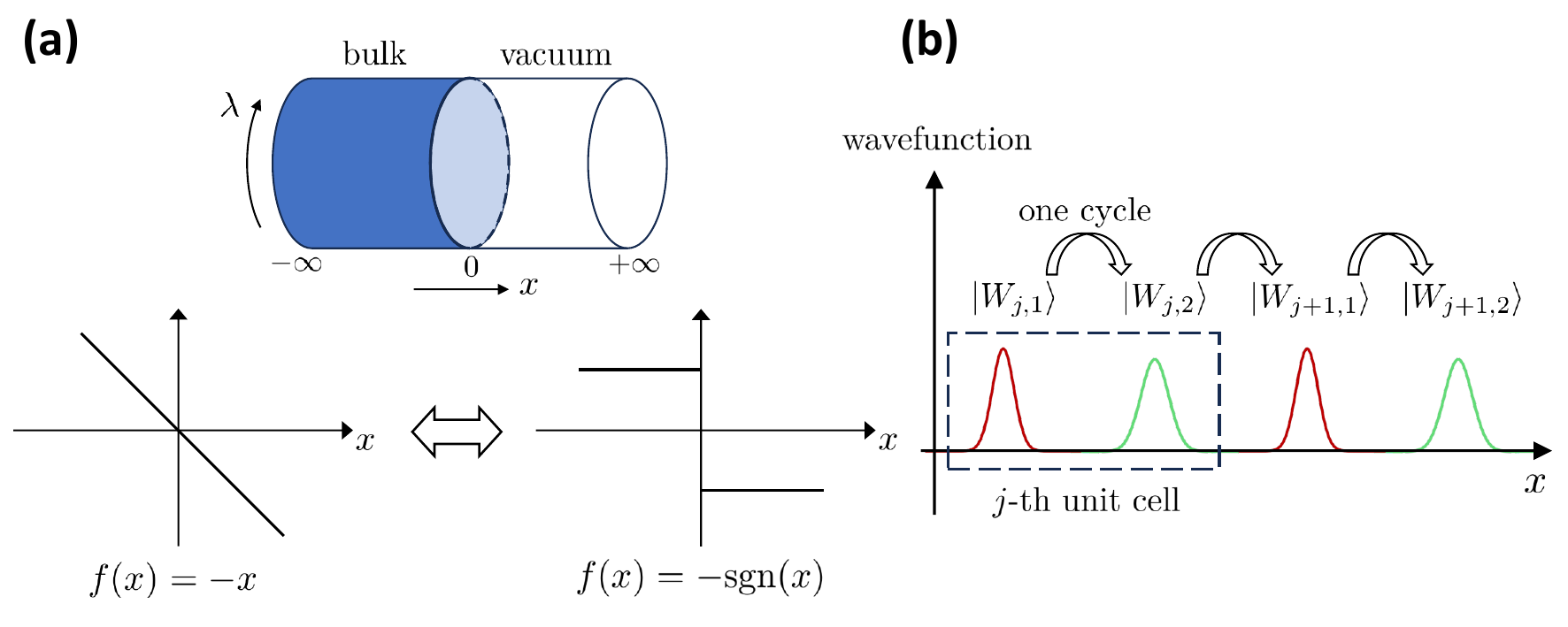}
	\caption{\label{Fig:bulk boundary correspondence}(a) The cylinder geometry. $\lambda$ direction is periodic, negative $x$ region is bulk and positive $x$ region is vacuum. $-PxP$ and flattened Hamiltonian $-P\mathrm{sgn}(x)P$ with open $x$ boundary are adiabatically connected by deforming $-x$ into $-\mathrm{sgn}(x)$. (b) Schematic of adiabatic evolution of MLWF in $n_b=2$, $C=1$ case.}
\end{figure}

Third, if $\{\phi_{\mu}\}$ are non-degenerate as $\lambda=0\rightarrow \lambda=2\pi$, we will find single MLWF evolves as
\begin{equation}\label{Eq:general WS state and eigenvalue evolution}
	\phi_{\mu}\rightarrow \phi_{1+(\mu+C-1)\mathrm{mod} n_b}+\lfloor(\mu+C-1)/n_b\rfloor, \ \ |{\cal W}^{\mathrm{MLWF}}_{j,\mu}\rangle \rightarrow |{\cal W}^{\mathrm{MLWF}}_{j+\lfloor (\mu+C-1)/n_b\rfloor,1+(\mu+C-1)\mathrm{mod} n_b}\rangle.
\end{equation}
As explained in the main text, the non-degeneracy is guaranteed by RSEG induced by Stark localization. We can also show the flow (\ref{Eq:general WS state and eigenvalue evolution}) with the help of bulk-boundary correspondence \cite{fidkowski2011model}. Consider the (1+1)-dimensional system in cylinder setup in Fig. \ref{Fig:bulk boundary correspondence}(a), which is half infinite in $x$ direction, and periodic in $\lambda$ direction. The MLWF center $\{\phi_{\mu}(\lambda)+j|j\in\mathbb{Z}\}$ in each one-dimensional subsystem with given $\lambda$ can be smoothly deformed to the edge mode spectrum $\{E_n(\lambda)\}$. This is because by flattening the dispersion, spectrum including edge mode can be recast as eigenvalues of $-P\mathrm{sgn}(x)P$ (we drop trivial eigenvalue $1$ from full Hamiltonian $-P\mathrm{sgn}(x)P+(1-P)$). Then we can deform $-\mathrm{sgn}(x)$ into $-x$ smoothly, so eigenvalues of $-P\mathrm{sgn}(x)P$ and $-PxP$ correspond adiabatically \cite{fidkowski2011model}. Given total Chern number $C$ of the $n_b$ bands, there will be $C$ chiral edge modes, which make $n$-th eigenstate $\left|\psi_n\right\rangle$ flows to $\left|\psi_{n-C}\right\rangle$ when $\lambda$ changes from $0$ to $2\pi$. Accordingly we know that MLWF $|{\cal W}^{\mathrm{MLWF}}\rangle$ flows to $|{\cal W}^{\mathrm{MLWF}}_{j+\lfloor (\mu+C-1)/n_b\rfloor,1+(\mu+C-1)\mathrm{mod} n_b}\rangle$. If $\{\phi_{\mu}\}$ were degenerate and do not flow as (\ref{Eq:general WS state and eigenvalue evolution}), then it would imply crossing between chiral edge states, which we do not expect to be true in general. Fig. \ref{Fig:bulk boundary correspondence}(b) is a schematic illustration of adiabatic evolution of MLWF, with $n_b=2$ and $C=1$. Also, in the single-band case, each MLWF moves by $C$ unit cells over one cycle, $|{\cal W}^{\mathrm{MLWF}}_{j}\rangle\rightarrow|{\cal W}^{\mathrm{MLWF}}_{j+C}\rangle$, thus contributing integer quantized charge transport (here we omit the $\mu$ index since there is only one MLWF per unit cell).

\begin{figure}[htbp]
	\includegraphics[width=0.6\columnwidth]{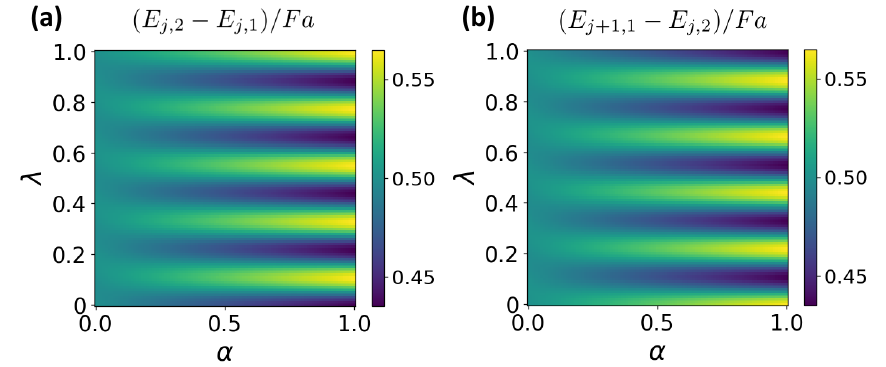}
	\caption{\label{Fig:adiabatic connection}Adiabatic connection between multiband Stark states at finite $\left|F\right|$ and maximally localized Wannier functions. The colormaps show gaps between adjacent multiband Stark states $\left|{\cal W}_{j,\mu}(\lambda,\alpha)\right\rangle$ (in units of $Fa$) which never close, ensuring adiabatic correspondence between MLWF $\left|{\cal W}^{\mathrm{MLWF}}_{j,\mu}(\lambda)\right\rangle$ and multiband Stark state $\left|{\cal W}_{j,\mu}(\lambda,\alpha=1)\right\rangle$. Due to adiabatic evolution (\ref{Eq:general WS state evolution}), the $\lambda=1$ part of (a) and $\lambda=0$ part of (b) are continuously connected, and vice versa.}
\end{figure}

Finally, by adiabatically connecting multiband Stark states at finite $\left|F\right|$
to MLWF, we obtain the adiabatic evolution (\ref{Eq:general WS state evolution}). In other words, we need to confirm RSEG never closes during $\alpha=0\rightarrow\alpha=1$ for Hamiltonian (\ref{Eq:interpolation Hamiltonian}), which is generally true in the absence of symmetry induced degeneracy. Here we exemplify with the $1/2$ fractional TSI based on $2/9$ GAA model in the main text, with $H_0(\lambda)=\sum_{m}\left[ J+\delta J\cos\left(\lambda-\frac{4\pi}{9}m\right) \right](c_{m+1}^{\dagger}c_{m}+\mathrm{h.c.}),\,J=-1,\,\delta J=-0.98$. We choose $Fa=0.072$, which is in the fractional phase for finite chain. As plotted Fig. \ref{Fig:adiabatic connection}, the RSEG $E_{j,2}-E_{j,1}$ and $E_{j+1,1}-E_{j,2}$ never close for all $(\lambda,\alpha)\in [0,2\pi]\times[0,1]$, thus validating the adiabatic connection, where $E_{j,\mu}(\lambda,\alpha)$ is the eigenenergy of multiband Stark state $\left|{\cal W}_{j,\mu}(\lambda,\alpha)\right\rangle$.

\section{Chern number in synthetic Brillouin zone}\label{SupSec:Chern number}

In this section we show the Chern number $C^{\prime}$ defined in the synthetic Brillouin zone is the same as total Chern number $C$ of the $n_b$ Chern bands. The Chern number associated with many-body state $\left|\Psi_{1}(k_{x,\alpha},k_{y,\alpha}=\lambda/n_b))\right\rangle= \left|{\cal V}_{k_x=0,1},\dots,{\cal V}_{k_x,1}\dots,{\cal V}_{k_x=2\pi,1}\right\rangle$ in synthetic Brilluoin zone is
\begin{equation}
\begin{split}
    C^{\prime}&=\frac{1}{2\pi}\int_{k_x=0}^{2\pi}\int_{k_y=0}^{2\pi}\tilde{\Omega}_{yx}\mathrm{d}k_x \mathrm{d}k_y
    =\frac{1}{2\pi}\int_{k_x=0}^{2\pi}\int_{\lambda=0}^{2\pi n_b}\tilde{\Omega}_{\lambda x}\mathrm{d}k_x \mathrm{d}\lambda
    =\frac{1}{2\pi}\int_{k_x=0}^{2\pi}\int_{\lambda=0}^{2\pi n_b}\partial_{\lambda}\tilde{A}_x \mathrm{d}k_x \mathrm{d}\lambda\\
    &=\frac{1}{2\pi}\int_{k_x=0}^{2\pi}\int_{\lambda=0}^{2\pi}\partial_{\lambda}\left[ \sum_{k=0}^{n_b-1} \tilde{A}_{x}(\lambda+2\pi k)\right] \mathrm{d}k_x \mathrm{d}\lambda,
\end{split}
\end{equation}
where $\tilde{\Omega}_{yx},\,\tilde{\Omega}_{\lambda x},\,\tilde{A}_x$ are Berry curvature and connection from $\left|{\cal V}_{k_x,k_y=\lambda/n_b,1}\right\rangle=\left|{\cal V}_{k_x,1}(\lambda)\right\rangle$,
\begin{equation}
    \tilde{\Omega}_{yx}=-2\mathrm{Im}\left\langle \partial_{k_y}{\cal V}_{k_x,k_y,1}|\partial_{k_x}{\cal V}_{k_x,k_y,1}\right \rangle,\,\tilde{\Omega}_{\lambda x}=-2\mathrm{Im}\left\langle \partial_{\lambda}{\cal V}_{k_x,1}(\lambda)|\partial_{k_x}{\cal V}_{k_x,1}(\lambda)\right \rangle,\,\tilde{A}_x=\mathrm{i}\left\langle {\cal V}_{k_x,1}(\lambda)|\partial_{k_x}{\cal V}_{k_x,1}(\lambda)\right \rangle,
\end{equation}
and in the second step we applied periodicity of Bloch states in $k_x$ direction. Because $\left\lbrace \left|{\cal V}_{k_x,1}(\lambda+2\pi k)\right\rangle\right|k=0,\dots,n_b-1\rbrace$ are orthogonal, they must form a complete orthonormal basis in the $n_b$-dimensional Hilbert space at fixed $(k_x,\lambda)$. Thus they are related to the Bloch states before we apply linear potential $| u^n_{k_x}(\lambda) \rangle$ by a unitary transformation $\mathbf{V}$,
\begin{equation}
	\left( \left|{\cal V}_{k_x,1}(\lambda)\right\rangle,\dots,\left|{\cal V}_{k_x,1}(\lambda+2\pi (n_b-1))\right\rangle \right)=\left( |u_{k_x}^{1}\rangle,\dots,|u_{k_x}^{n_b}\rangle \right) \mathbf{V}.
\end{equation}
Correspondingly, the Berry connection transforms as
\begin{equation}
	\sum_{k=0}^{n_b-1} \tilde{A}_{x}(\lambda+2\pi k)=\mathrm{tr}(\mathbf{V}^{\dagger}\mathbf{A}_{x}\mathbf{V})+\mathrm{tr}(\mathrm{i}\mathbf{V}^{\dagger}\partial_{k_x}\mathbf{V}))=\mathrm{tr}(\mathbf{A}_{x})+\mathrm{tr}(\mathrm{i}\mathbf{V}^{\dagger}\partial_{k_x}\mathbf{V}))=\sum_{n}\mathbf{A}_{x}^{nn}+\mathrm{tr}(\mathrm{i}\mathbf{V}^{\dagger}\partial_{k_x}\mathbf{V}).
\end{equation}
The Chern number $C^{\prime}$ is translated into total Chern number $C$ plus a vanishing residual term,
\begin{equation}
	\begin{split}
		C^{\prime}&=\frac{1}{2\pi}\int_{k_x=0}^{2\pi}\int_{\lambda=0}^{2\pi}\partial_{\lambda}\left[ \sum_{k=0}^{n_b-1} \tilde{A}_{x}(\lambda+2\pi k)\right]\mathrm{d}k_x\mathrm{d}\lambda \\
		&=\frac{1}{2\pi}\int_{k_x=0}^{2\pi}\int_{\lambda=0}^{2\pi}\sum_{n}\partial_{\lambda}\mathbf{A}_{x}^{nn}\mathrm{d}k_x\mathrm{d}\lambda
		+\frac{1}{2\pi}\int_{\lambda=0}^{2\pi} \partial_{\lambda}\left[\int_{k_x=0}^{2\pi} \mathrm{tr}(\mathrm{i}\mathbf{V}^{\dagger}\partial_{k_x}\mathbf{V})\mathrm{d}k_x\right]\mathrm{d}\lambda\\
		&=\frac{1}{2\pi}\int_{k_x=0}^{2\pi}\int_{\lambda=0}^{2\pi}\sum_{n}\partial_{\lambda}\mathbf{A}_{x}^{nn}\mathrm{d}k_x\mathrm{d}\lambda \\
		&=\frac{1}{2\pi}\int_{k_x=0}^{2\pi}\int_{\lambda=0}^{2\pi}\sum_{n}\Omega_{\lambda x}^{n}\mathrm{d}k_x\mathrm{d}\lambda \\
		&=\sum_{n}C_n=C,
	\end{split}
\end{equation}
We prove the residual term is zero as follows. Note that effect of infinitesimal step in $k_x$ acts on $\mathbf{V}$ as
\begin{equation}
	\mathbf{V}(k_x)(\mathbf{I}+\mathbf{V}^{\dagger}\partial_{k_x}\mathbf{V}\Delta k_x)=\mathbf{V}(k_x+\Delta k_x)+o(\Delta k_x).
\end{equation}
Then we can integrate steps to obtain
\begin{equation}
	\mathbf{V}(k_x=0)\mathcal{P}\exp\left[ \int_{k_x=0}^{2\pi} \mathbf{V}^{\dagger}\partial_{k_x}\mathbf{V}\mathrm{d}k_x\right]=\mathbf{V}(k_x=2\pi),
\end{equation}
where $\mathcal{P}$ indicates path order. Recall the periodicity in $k_x$ direction, we have
\begin{equation}
	\mathbf{V}(k_x=0)=\mathbf{V}(k_x=2\pi),\,\mathcal{P}\exp\left[ \int_{k_x=0}^{2\pi} \mathbf{V}^{\dagger}\partial_{k_x}\mathbf{V}\mathrm{d}k_x\right]=\mathbf{I}.
\end{equation}
Finally
\begin{equation}
	\exp\left[ \int_{k_x=0}^{2\pi} \mathrm{tr}(\mathbf{V}^{\dagger}\partial_{k_x}\mathbf{V})\mathrm{d}k_x\right] =\det\left\lbrace \mathcal{P}\exp\left[ \int_{k_x=0}^{2\pi} \mathbf{V}^{\dagger}\partial_{k_x}\mathbf{V}\mathrm{d}k_x\right] \right\rbrace = \det\mathbf{I} = 1,\,\int_{k_x=0}^{2\pi} \mathrm{tr}(\mathrm{i}\mathbf{V}^{\dagger}\partial_{k_x}\mathbf{V})\mathrm{d}k_x=2k\pi,\,k\in\mathbb{Z}.
\end{equation}
Because $\int_{k_x=0}^{2\pi} \mathrm{tr}(\mathrm{i}\mathbf{V}^{\dagger}\partial_{k_x}\mathbf{V})\mathrm{d}k_x$ is quantized and smooth in $\lambda$, its derivative must vanish,
\begin{equation}
	\partial_{\lambda}\left[\int_{k_x=0}^{2\pi} \mathrm{tr}(\mathrm{i}\mathbf{V}^{\dagger}\partial_{k_x}\mathbf{V})\mathrm{d}k_x\right]=0.
\end{equation}
Thus we complete the proof that $C^{\prime}=C$.

\section{Real space energy gap scaling near transition point}\label{SupSec:RSEG scaling}

In this section, we derive the scaling of real space energy gap (RSEG) $\Delta^{\mathrm{RS}}$ near the transition point $\left|F\right|=0$. As described in the main text, RSEG originates in the hybridization of two degenerate Stark states $|w_{j_1}^{(1)}\rangle$ $(|w_{j_2}^{(2)}\rangle)$ from lower (upper) band located at $x_1$ and $x_2$, with their separation satisfying $\Delta x_{1,2}=\left|x_1-x_2\right|=(\overline{E}_2-\overline{E}_1)/\left|F\right|$. In the limit $\left|F\right|\rightarrow 0$, we have $\Delta x_{1,2}\gg a$ that the two coupled Stark states are far apart in real space. It follows that when computing the RSEG or coupling matrix element between these two states, only long-range asymptotic behavior of the Stark states needs to be considered, giving the asymptotic expression
\begin{equation}
    \begin{split}
        w_{j_1}^{(1)}(x)&\sim P^{(1)}\tilde{w}_{j_1}^{(1)}(x) \sim P^{(1)}\frac{1}{\sqrt{\xi^{(1)}}}\exp\left(-\left|x-x_{1}\right|/\xi^{(1)}\right),\\
        w_{j_2}^{(2)}(x)&\sim P^{(2)}\tilde{w}_{j_2}^{(2)}(x) \sim P^{(2)}\frac{1}{\sqrt{\xi^{(2)}}}\exp\left(-\left|x-x_{2}\right|/\xi^{(2)}\right),\\
        \xi^{(1,2)}&\sim \frac{\sqrt{2}W_b^{(1,2)}}{\left|F\right|},
    \end{split}
\end{equation}
with band projections $P^{(1,2)}$, localization lengths $\xi^{(1,2)}$ and band widths $W_b^{(1,2)}$. Here the exponential part describes localization of the wavefunction among unit cells, and band projections $P^{(1,2)}$ dress the sublattice degree of freedom. Our goal is the coupling matrix element $\langle w_{j_1}^{(1)}|F\hat{x}|w_{j_2}^{(2)}\rangle=F\langle\tilde{w}_{j_1}^{(1)}|P^{(1)}\hat{x}P^{(2)}|\tilde{w}_{j_2}^{(2)}\rangle$. Specifically, the operator $P^{(1)}\hat{x}P^{(2)}$ commutes with lattice translation, which results in the following two properties. First, $P^{(1)}\hat{x}P^{(2)}$ is bounded such that $\|P^{(1)}\hat{x}P^{(2)}|\tilde{w}_{j_2}^{(2)}\rangle\|<l_0\||\tilde{w}_{j_2}^{(2)}\rangle\|$, where $l_0<\infty$ and is independent of $F$ \cite{marzari1997maximally}. This is in sharp contrast to operator $\hat{x}$, which is unbounded in the sense that $\|\hat{x}|\tilde{w}_{j_2}^{(2)}\rangle\|$ diverges if $|\tilde{w}_{j_2}^{(2)}\rangle$ is located at $\left|x_2\right|\rightarrow\infty$. Second, applying $P^{(1)}\hat{x}P^{(2)}$ does not affect the exponential decay of wavefunction across unit cells, so $P^{(1)}\hat{x}P^{(2)}|\tilde{w}_{j_2}^{(2)}\rangle$ is also a exponentially localized state with localization length $\xi^{(2)}$. The exponential localization of $|\tilde{w}_{j_2}^{(2)}\rangle$ implies $\mathcal{T}(a)\tilde{w}_{j_2}^{(2)}(x)=\tilde{w}_{j_2}^{(2)}(x+a)\sim\mathrm{e}^{-a/\xi^{(2)}}\tilde{w}_{j_2}^{(2)}(x+a)$ for $x>x_2$ ($\mathcal{T}(a)$ is translation by one unit cell), and by multiplying $P^{(1)}\hat{x}P^{(2)}$ to the left we obtain $P^{(1)}\hat{x}P^{(2)}\mathcal{T}(a)\tilde{w}_{j_2}^{(2)}(x)=\mathcal{T}(a)P^{(1)}\hat{x}P^{(2)}\tilde{w}_{j_2}^{(2)}(x)=P^{(1)}\hat{x}P^{(2)}\tilde{w}_{j_2}^{(2)}(x+a)\sim\mathrm{e}^{-a/\xi^{(2)}}P^{(1)}\hat{x}P^{(2)}\tilde{w}_{j_2}^{(2)}(x+a)$, implying the exponential localization. Combining these two properties, we know that the magnitude of the coupling matrix element $F\langle\tilde{w}_{j_1}^{(1)}|P^{(1)}\hat{x}P^{(2)}|\tilde{w}_{j_2}^{(2)}\rangle$ can be estimated from that of the overlap integral $F\langle\tilde{w}_{j_1}^{(1)}|\tilde{w}_{j_2}^{(2)}\rangle$, which is
\begin{equation}
    \begin{split}                 &F\langle\tilde{w}_{j_1}^{(1)}|\tilde{w}_{j_2}^{(2)}\rangle=F\int \tilde{w}_{j_1}^{(1)*}(x)\tilde{w}_{j_2}^{(2)}(x) \mathrm{d}x\\
    &\sim \frac{F}{\sqrt{\xi^{(1)}\xi^{(2)}}}\int \exp\left(-\frac{\left|x-x_{1}\right|}{\xi^{(1)}}-\frac{\left|x-x_{2}\right|}{\xi^{(2)}}\right)\mathrm{d}x\\
    &=\frac{F}{\sqrt{\xi^{(1)}\xi^{(2)}}}\left\{\frac{\xi^{(1)}\xi^{(2)}}{\xi^{(1)}+\xi^{(2)}}\exp\left(-\frac{\Delta x_{1,2}}{\xi^{(1)}}-\frac{\Delta x_{1,2}}{\xi^{(2)}}\right)+\frac{\xi^{(1)}\xi^{(2)}}{\xi^{(1)}-\xi^{(2)}}\left[\exp\left(-\frac{\Delta x_{1,2}}{\xi^{(2)}}\right)-\exp\left(-\frac{\Delta x_{1,2}}{\xi^{(1)}}\right)\right]\right\}\\
    &\sim F\exp\left(-\frac{\Delta x_{1,2}}{\xi^{(1,2)}}\right)\\
    &=O(\left|F\right|).
    \end{split}
\end{equation}
In the last step, we used $\xi^{(1,2)},\Delta x_{1,2}=O(\left|F\right|^{-1})$, rendering the exponential factor trivial. In other words, the separation between degenerate Stark states $\Delta x_{1,2}$ and their spatial extent $\xi^{(1,2)}$ increase equally fast and cancel as $\left|F\right|\rightarrow0$. Finally, we obtain the RSEG
\begin{equation}
    \Delta^{\mathrm{RS}}=2\left|\langle w_{j_1}^{(1)}|F\hat{x}|w_{j_2}^{(2)}\rangle\right|\sim 2\left|F\langle\tilde{w}_{j_1}^{(1)}|\tilde{w}_{j_2}^{(2)}\rangle\right|\sim 2\left|F\right|\exp\left(-\frac{\Delta x_{1,2}}{\xi^{(1,2)}}\right)=O(\left|F\right|).
\end{equation}

\section{Fractionally quantized charge pumping}\label{SupSec:fractional pump}

In this section we show that the topology of the fractional topological Stark insulator can be probed through fractionally quantized charge pumping, that transferred charge is topologically quantized only for multiple drive cycles. As in Fig. \ref{Fig:fractional topology}, the $n_b$ cycle adiabatic charge transport of the fractional topological Stark insulator is the same as one cycle Thouless pumping of the mapped topological band insulator which exhibits topological quantization. This is because charge polarization is determined solely by the insulating wavefunction, and adiabatic charge transport can be derived from the change in polarization \cite{resta1992theory,king1993theory,resta1994macroscopic}. The physical validity of this mapping is ensured by the existence of a non-interacting lattice-periodic Hamiltonian $H^{\prime}(\lambda)$ with periodicity $2\pi n_b$, which hosts one fractional topological Stark insulating state $\left|\Psi_1(k_{x,\alpha},\lambda)\right\rangle$ as a band insulator and generates the same adiabatic evolution as $H(\lambda)$. A natural approach to construct $H^{\prime}(\lambda)$ is to require artificial Bloch states $\left|{\cal V}_{k_x,1}(\lambda)\right\rangle$ at all $k_x\in[0,2\pi]$ to form a gapped band, with one possible form being $H^{\prime}(\lambda)=-E_0\sum_{k_x}\left|{\cal V}_{k_x,1}(\lambda)\right\rangle\left\langle{\cal V}_{k_x,1}(\lambda)\right|,\,E_0>0$. It follows that $\left|\Psi_1(k_{x,\alpha},\lambda)\right\rangle$ is a band insulator under $H^{\prime}(\lambda)$ by filling a gapped band, and the constructed $H^{\prime}(\lambda)$ has the same periodicity $2\pi n_b$ in $\lambda$ as $\left|{\cal V}_{k_x,1}(\lambda)\right\rangle$, consistent with the mapping. Through this construction, the $n_b$ cycle charge pumping of $\left|\Psi_1(k_{x,\alpha},\lambda)\right\rangle$ is mapped to one cycle of Thouless pumping under $H^{\prime}(\lambda)$, so the pumped charge must be topologically quantized to the Chern number in the synthetic Brillouin zone $C^{\prime}=C$. Thus we obtain $C/n_b$ fractionally quantized charge pumping in which transferred charge is quantized to $C$ over $n_b$ cycles, which can be utilized to probe fractional topology.

\section{The $1/3$ fractional topological Stark insulator}\label{SupSec:1/3 pumping}

\begin{figure}[htbp]
	\includegraphics[width=\columnwidth]{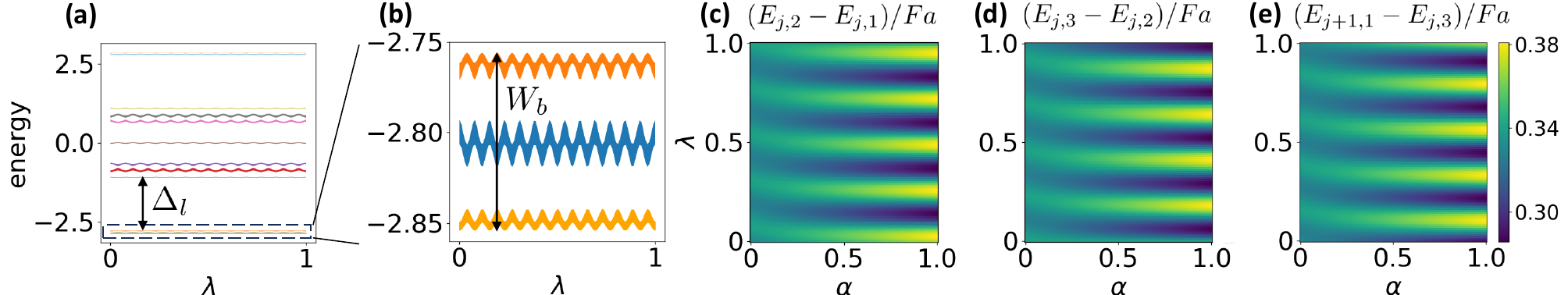}
	\caption{\label{Fig:band structure 3/13}(a-b) Band structure of $3/13$ AA model with respect to parameter $\lambda$. Color is filled within energy range of each one-dimensional band. Here first three bands have width $W_b=0.098$, and are separated from higher bands by large gap $\Delta_l=1.65$. (c-e) Adiabatic connection between multiband Stark states and MLWF. As indicated in colormaps, energy gaps between adjacent multiband Stark states $\left|{\cal W}_{j,\mu}(\lambda,\alpha)\right\rangle$ (in units of $Fa$) never close, guaranteeing adiabatic correspondence between MLWF $\left|{\cal W}^{\mathrm{MLWF}}_{j,\mu}(\lambda)\right\rangle$ and multiband Stark state $\left|{\cal W}_{j,\mu}(\lambda)\right\rangle$. Similar to Fig. \ref{Fig:adiabatic connection}, the $\lambda=1$ part of (c) is continuously connected to $\lambda=0$ part of (d), and so on, according to Eq. (\ref{Eq:3/13 WS pump}).}
\end{figure}

In this section, we further provide an example of $1/3$ fractional TSI to show the generality of our theory. This is based on a $3/13$ commensurate Aubry-Andr\'{e} (AA) model with modulated on-site energy
\begin{equation}\label{Eq:3/13 model}
	H_{0}(\lambda)=\sum_{m}J(c_{m+1}^{\dagger}c_{m}+\mathrm{h.c.})+\sum_{m}V\cos\left (\lambda-\frac{6\pi}{13}m\right) c_{m}^{\dagger}c_{m},
\end{equation}
where we choose $J=-1,\,V=-2$. The band structure with versus pump parameter $\lambda$ is plotted in Fig. \ref{Fig:band structure 3/13}(a-b), where first $n_b=3$ bands are isolated and flat, with band width $W_b=0.098$ and separated from fourth band by a large gap $\Delta_l=1.65$. Moreover, the first three bands carry Chern number $C_1=-4,\,C_2=9,\,C_3=-4$ that sum up to total Chern number $C=1$. According to the general theory, the fractional TSI phase is characterized by $C/n_{b}=1/3$ fractional number. The linear potential gradient is chosen to be $Fa=1.3$, satisfying $\Delta> Fa\gg W_b$, which is sufficient to enter the fractional regime for a finite chain. Similar to \ref{Supsec:multiband stark states}, we introduce interpolation Hamiltonian for adiabatic connection,
\begin{equation}
	\tilde{H}(\lambda,\alpha)=P(F\hat{x}+\alpha H_0(\lambda))P,
\end{equation}
together with multiband Stark states $\left|{\cal W}_{j,\mu}(\lambda,\alpha)\right\rangle$ and eigenvalues $E_{j,\mu}(\lambda,\alpha)$, labelled by unit cell index $j$ and species index $\mu=1,2,3$. We see from Fig. \ref{Fig:band structure 3/13}(c-e) that different multiband Stark states $\left|{\cal W}_{j,\mu}(\lambda,\alpha)\right\rangle$ are separated by finite RSEG. Thus multiband Stark states $\left|{\cal W}_{j,\mu}\right\rangle$ adiabatically evolve as
\begin{equation}\label{Eq:3/13 WS pump}
	\left|{\cal W}_{j,1}\right\rangle\rightarrow\left|{\cal W}_{j,2}\right\rangle,\,\left|{\cal W}_{j,2}\right\rangle\rightarrow\left|{\cal W}_{j,3}\right\rangle,\,\left|{\cal W}_{j,3}\right\rangle\rightarrow\left|{\cal W}_{j+1,1}\right\rangle,
\end{equation}
over one drive cycle $\lambda=0\rightarrow \lambda=2\pi$.

\begin{figure}[htbp]
	\includegraphics[width=0.7\columnwidth]{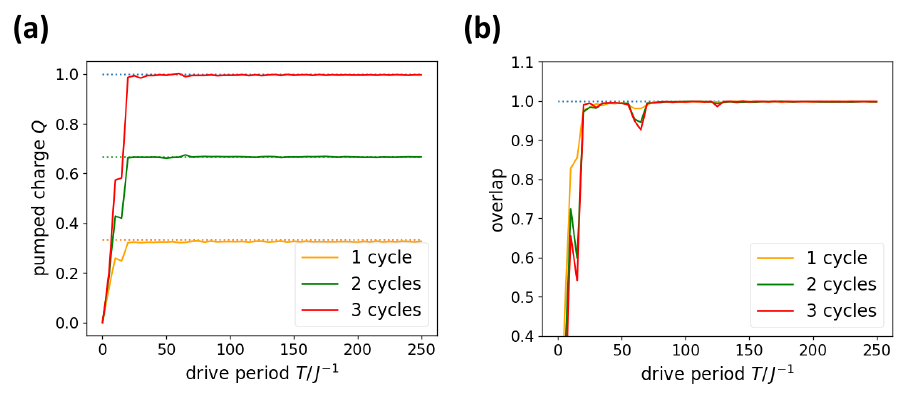}
	\caption{\label{Fig:fractional pump 3/13}(a) Pumped charge the fractional topological Stark insulating state. In the three-cycle quantized plateau for drive period $T>100 J^{-1}$, pumped charge $Q\in[0.994,0.998]$. (b) Overlaps between evolved wavefunction and instantaneous multiband Stark states. In the plateau, the overlaps lie in $[0.996,1.0]$.}
\end{figure}

In Fig. \ref{Fig:fractional pump 3/13} we plot the results of time evolution for the $1/3$ fractional TSI under full Hamiltonian $H(\lambda)$. Fig. \ref{Fig:fractional pump 3/13}(a) shows that the three-cycle pumped charge is quantized to $C=1$ for drive period $T>100 J^{-1}$, while for one or two cycles, the pumped charge is fractionalized to be $1/3$ and $2/3$, respectively, agreeing with the $1/3$ fractional charge pumping. Fig. \ref{Fig:fractional pump 3/13}(b) shows overlaps $|\left\langle{\cal W}_{j,2}\right|U(T)\left|{\cal W}_{j,1}\right\rangle|^2$, $|\left\langle{\cal W}_{j,3}\right|U(2T)\left|{\cal W}_{j,1}\right\rangle|^2$ and $|\left\langle{\cal W}_{j+1,1}\right|U(3T)\left|{\cal W}_{j,1}\right\rangle|^2$, which has plateau close to $1$, confirming adiabatic evolution Eq. (\ref{Eq:3/13 WS pump}).

\section{Parameters for the experimental scheme}\label{SupSec:experimental parameter}

In this section we detail the derivation of tight-binding Hamiltonian and estimation of its parameters. In the tight-binding limit, the basis is set to be maximally localized Wannier states of the lowest band of the primary lattice (wavelength $\lambda_1$) $\left| w^{\mathrm{OL}}_m \right\rangle$, labelled by lattice site $m$. Retaining only the nearest-neighbor hopping term, and on-site energy term from secondary lattice, we obtain the Hamiltonian matrix element as \cite{modugno2009exponential}
\begin{equation}
    \begin{split}
    \left\langle w^{\mathrm{OL}}_m \right| H^{\mathrm{OL}} \left| w^{\mathrm{OL}}_{m^{\prime}} \right\rangle &= J\delta_{m,m^{\prime}\pm 1} + V\cos\left( \lambda-2\pi\beta m\right)\delta_{m,m^{\prime}},\,\beta=\lambda_1/\lambda_2,\\
    J&=-\int w_{m+1}^{\mathrm{OL}}(x)^*\left[ \frac{\hat{p}_x^2}{2m_a} + V_1 \sin^2 \left( k_1 x \right)\right] w_{m}^{\mathrm{OL}}(x) \mathrm{d}x,\\
    V&=\frac{V_2 \beta^2}{2E_{R,2}}\int \cos(2\beta x)\left|w_{m}^{\mathrm{OL}}(x)\right|^2 \mathrm{d}x.
    \end{split}
\end{equation}
Therefore the tight-binding Hamiltonian reads
\begin{equation}
    H_{0}(\lambda)=\sum_{m}J(c_{m+1}^{\dagger}c_{m}+\mathrm{h.c.})+\sum_{m}V\cos\left( \lambda-2\pi\beta m\right) c_{m}^{\dagger}c_{m}.
\end{equation}
For the red detuned optical lattice here, the decay rate $\Gamma_{\mathrm{eff}}$ (lifetime $\tau=1/\Gamma_{\mathrm{eff}}$) is estimated to be \cite{jaksch2005cold}
\begin{equation}
    \Gamma_{\mathrm{eff}}\simeq \frac{\Gamma V_1}{\delta^{\mathrm{OL}}},
\end{equation}
where $\Gamma$ is decay rate of the excited state, and $\delta^{\mathrm{OL}}$ is the detuning of laser of wavelength $\lambda_1$. The contribution of secondary lattice is dropped since it is relatively very weak.

\section{Observing fractional charge pumping for generic initialization of the many-body state}\label{SupSec:initial state}

In this section we show that fractionally quantized charge pumping can be observed for more general initial states under proper drive period. Consider a many-body state $\left|\Psi\right\rangle=\left|\psi_1,\psi_2,\dots,\psi_j,\dots\right\rangle$ obtained by occupying a general single-particle localized state $\left|\psi_{j}\right\rangle$ in every unit cell, with $\left|\psi_{j}\right\rangle$ restricted to the $n_b$ bands subspace and $j$ being unit cell label. If drive period $T$ satisfies $T=kT_{\mathrm{Bloch}},\,k\in\mathbb{Z}$ with Bloch oscillation period $T_{\mathrm{Bloch}}=2\pi/Fa$, then transferred charge will be quantized to $C$ over $n_b$ cycles. This is because the dynamical phases of eigenstates $\left|{\cal W}_{j,\mu}\right\rangle$ differ by integer multiples of $2\pi$, that the superposition state $\left|\psi_{j}\right\rangle$ is shifted by $C$ unit cells over $n_b$ cycles and does not dephase. It is important to note that $\left|\Psi\right\rangle$ only serves as a probe of the fractional TSI, but lacks the topological robustness against perturbation (disorder for example) of the exact fractional TSI state.

We start by noting that multiband Stark states $\left|{\cal W}_{j,\mu}\right\rangle$ form a complete and orthogonal basis in the $n_b$ band subspace, so we can expand any single-particle localized state at unit cell $j$ in this subspace as
\begin{equation}
	\left|\psi_j\right\rangle=\sum_{l,\mu}c_{l,\mu}\left|{\cal W}_{l,\mu}\right\rangle,
\end{equation}
where $c_{l,\mu}$ are expansion coefficients. After one adiabatic cycle, the evolved state is
\begin{equation}
	U(T)\left|\psi_j\right\rangle=\sum_{l,\mu}c_{l,\mu}\mathrm{e}^{-\mathrm{i}\theta_{l,\mu}(T)}\left|{\cal W}_{l+\lfloor (\mu+C-1)/n_b\rfloor,1+(\mu+C-1)\mathrm{mod} n_b}\right\rangle,
\end{equation}
where we denote the dynamical phase for $\left|{\cal W}_{l,\mu}\right\rangle$ initial state as $\theta_{l,\mu}(T)$. Because $\left|{\cal W}_{l,\mu}\right\rangle$ are not degenerate, they pick up different dynamical phases $\theta_{l,\mu}(T)$ and the superposition state dephases. Even after $n_b$ cycles, the evolved state
\begin{equation}\label{Eq:nb cycle state}
	U(n_b T)\left|\psi_j\right\rangle=\sum_{l,\mu}c_{l,\mu}\mathrm{e}^{-\mathrm{i}\theta_{l,\mu}(n_b T)}\left|{\cal W}_{l+C,\mu}\right\rangle,
\end{equation}
is not the initial state translated by $C$ unit cells. Also the change of position expectation value deviates from $Ca$,
\begin{equation}\label{Eq:nb cycle displacement}
	\begin{split}
		\Delta x(n_b T)=&\left\langle \psi_j\right| U(n_b T)^{\dagger}\hat{x}U(n_b T) \left|\psi_j\right\rangle - \left\langle \psi_j\right|\hat{x} \left|\psi_j\right\rangle \\
		=&\sum_{l,l^{\prime},\mu,\mu^{\prime}}c_{l,\mu}^{*}c_{l^{\prime},\mu^{\prime}}\left[ \mathrm{e}^{-\mathrm{i}(\theta_{l^{\prime},\mu^{\prime}}(n_b T)-\theta_{l,\mu}(n_b T))}\left\langle {\cal W}_{l+C,\mu}\right|\hat{x} \left|{\cal W}_{l^{\prime}+C,\mu^{\prime}}\right\rangle - \left\langle {\cal W}_{l,\mu}\right|\hat{x} \left|{\cal W}_{l^{\prime},\mu^{\prime}}\right\rangle \right] \\
		=&\sum_{l,\mu}|c_{l,\mu}|^2 \left[ \left\langle {\cal W}_{l+C,\mu}\right|\hat{x} \left|{\cal W}_{l+C,\mu}\right\rangle - \left\langle {\cal W}_{l,\mu}\right|\hat{x} \left|{\cal W}_{l,\mu}\right\rangle \right] \\
		& + \sum_{l,\mu\neq l^{\prime},\mu^{\prime}}c_{l,\mu}^{*}c_{l^{\prime},\mu^{\prime}}\left[ \mathrm{e}^{-\mathrm{i}(\theta_{l^{\prime},\mu^{\prime}}(n_b T)-\theta_{l,\mu}(n_b T))}\left\langle {\cal W}_{l+C,\mu}\right|\hat{x} \left|{\cal W}_{l^{\prime}+C,\mu^{\prime}}\right\rangle - \left\langle {\cal W}_{l,\mu}\right|\hat{x} \left|{\cal W}_{l^{\prime},\mu^{\prime}}\right\rangle \right] \\
		=& Ca\sum_{l,\mu}|c_{l,\mu}|^2 + \sum_{l,\mu\neq l^{\prime},\mu^{\prime}}c_{l,\mu}^{*}c_{l^{\prime},\mu^{\prime}}\left[ \mathrm{e}^{-\mathrm{i}(\theta_{l^{\prime},\mu^{\prime}}(n_b T)-\theta_{l,\mu}(n_b T))}\left\langle {\cal W}_{l,\mu}\right|\hat{x}+Ca \left|{\cal W}_{l^{\prime},\mu^{\prime}}\right\rangle - \left\langle {\cal W}_{l,\mu}\right|\hat{x} \left|{\cal W}_{l^{\prime},\mu^{\prime}}\right\rangle \right] \\
		=& Ca + \sum_{l,\mu\neq l^{\prime},\mu^{\prime}}c_{l,\mu}^{*}c_{l^{\prime},\mu^{\prime}}\left[ \left(\mathrm{e}^{-\mathrm{i}(\theta_{l^{\prime},\mu^{\prime}}(n_b T)-\theta_{l,\mu}(n_b T))}-1 \right)\left\langle {\cal W}_{l,\mu}\right|P\hat{x}P \left|{\cal W}_{l^{\prime},\mu^{\prime}}\right\rangle\right].
	\end{split}
\end{equation}
Since $\left|{\cal W}_{l,\mu}\right\rangle$ are not eigenstates of $P\hat{x}P$, they are not orthogonal with respect to $P\hat{x}P$, and the off-diagonal term cannot cancel.

However, if we choose specific driving period, the dynamical phases can be cancelled. First, Stark states of same kind $\left|{\cal W}_{l,\mu}\right\rangle$ are related by unit cell translation and their energy differ by linear potential difference, $E_{l^{\prime},\mu}-E_{l,\mu}=(l^{\prime}-l)Fa$. Consequently, their dynamical phases must satisfy
\begin{equation}
	\theta_{l^{\prime},\mu}(n_b T)-\theta_{l,\mu}(n_b T)=\int_{0}^{n_b T}[E_{l^{\prime},\mu}(\lambda)-E_{l,\mu}(\lambda)]\mathrm{d}t=\int_{0}^{n_b T}(l^{\prime}-l)Fa\mathrm{d}t=n_b T (l^{\prime}-l)Fa.
\end{equation}
If we choose $T$ such that $n_b T (l^{\prime}-l)Fa=2k\pi$ for all $l^{\prime}-l$, or $T=2k\pi/n_b Fa,\,k\in\mathbb{Z}$, then such dynamical phases will cancel, $\mathrm{e}^{-\mathrm{i}(\theta_{l^{\prime},\mu}(n_b T)-\theta_{l,\mu}(n_b T))}=1$. Second, for Stark states of different kinds, say $\left|{\cal W}_{l,\mu}\right\rangle$ and $\left|{\cal W}_{l,\mu^{\prime}}\right\rangle,\,\mu^{\prime}\neq\mu$, they are not related by unit cell translation. Nevertheless, we can show the dynamical phases cancel for driving period $T=2k\pi/Fa$. A simple interpretation is that one will traverse all $n_b$ species of Stark states over $n_b$ cycles starting from any $\left|{\cal W}_{l,\mu}\right\rangle$. Due to fractional pumping, we have $U(T)\left|{\cal W}_{l,\mu}\right\rangle\propto\left|{\cal W}_{l+\lfloor (\mu+C-1)/n_b\rfloor,1+(\mu+C-1)\mathrm{mod} n_b}\right\rangle,\,U(2T)\left|{\cal W}_{l,\mu}\right\rangle\propto\left|{\cal W}_{l+\lfloor (\mu+2C-1)/n_b\rfloor,1+(\mu+2C-1)\mathrm{mod} n_b}\right\rangle,\dots,U(n_b T)\left|{\cal W}_{l,\mu}\right\rangle\propto\left|{\cal W}_{l+C,\mu}\right\rangle$. Because $C$ and $n_b$ are coprime integers, $\{\mu,1+(\mu+C-1)\mathrm{mod} n_b,1+(\mu+2C-1)\mathrm{mod} n_b,\dots,1+(\mu+(n_b-1)C-1)\mathrm{mod} n_b\}=\{1,2,\dots,n_b\}$. Then there must be some integer $n_{\mu\mu^{\prime}}\in\{1,2,\dots,n_b-1\}$ such that $1+(\mu+n_{\mu\mu^{\prime}}C-1)\mathrm{mod} n_b=\mu^{\prime}$. The dynamical phases can be rewritten as
\begin{equation}
	\begin{split}
		\theta_{l,\mu}(n_b T)-\theta_{l,\mu^{\prime}}(n_b T)=&\int_{0}^{n_b T}[E_{l,\mu}(\lambda)-E_{l,\mu^{\prime}}(\lambda)]\mathrm{d}t \\
		=&\int_{0}^{n_{\mu\mu^{\prime}} T} E_{l,\mu}(\lambda)\mathrm{d}t+\int_{n_{\mu\mu^{\prime}}T}^{ n_b T} E_{l,\mu}(\lambda)\mathrm{d}t-\int_{0}^{n_b T} E_{l,\mu^{\prime}}(\lambda)\mathrm{d}t \\
		=&\int_{0}^{n_{\mu\mu^{\prime}} T} [E_{l,\mu^{\prime}}(\lambda+n_b-n_{\mu\mu^{\prime}})-CFa]\mathrm{d}t+\int_{n_{\mu\mu^{\prime}}T}^{ n_b T}[E_{l,\mu^{\prime}}(\lambda-n_{\mu\mu^{\prime}})+\lfloor (\mu+n_{\mu\mu^{\prime}}C-1)/n_b\rfloor Fa]\mathrm{d}t \\
		&-\int_{0}^{n_b T} E_{l,\mu^{\prime}}(\lambda)\mathrm{d}t \\
		=&\int_{(n_b-n_{\mu\mu^{\prime}})T}^{n_b T} [E_{l,\mu^{\prime}}(\lambda)-CFa]\mathrm{d}t+\int_{0}^{(n_b-n_{\mu\mu^{\prime}})T} [E_{l,\mu^{\prime}}(\lambda)+\lfloor (\mu+n_{\mu\mu^{\prime}}C-1)/n_b\rfloor Fa]\mathrm{d}t  \\
		&-\int_{0}^{n_b T} E_{l,\mu^{\prime}}(\lambda)\mathrm{d}t \\
		=&T\left[ (n_b-n_{\mu\mu^{\prime}})\lfloor (\mu+n_{\mu\mu^{\prime}}C-1)/n_b\rfloor-n_{\mu\mu^{\prime}}C\right] Fa.
	\end{split}
\end{equation}
By choosing $T=2k\pi/Fa$, the second kind of dynamical phases can be cancelled, $\mathrm{e}^{-\mathrm{i}(\theta_{l,\mu}(n_b T)-\theta_{l,\mu^{\prime}}(n_b T))}=1$. It follows that general dynamical phases cancel, by decomposition $\mathrm{e}^{-\mathrm{i}(\theta_{l,\mu}(n_b T)-\theta_{l^{\prime},\mu^{\prime}}(n_b T))}=\mathrm{e}^{-\mathrm{i}(\theta_{l,\mu}(n_b T)-\theta_{l,\mu^{\prime}}(n_b T))}\mathrm{e}^{-\mathrm{i}(\theta_{l,\mu^{\prime}}(n_b T)-\theta_{l^{\prime},\mu^{\prime}}(n_b T))}=1$.

In short, by setting drive period as integer multiples of Bloch oscillation period $T=2k\pi/Fa=kT_{\mathrm{Bloch}}$, the dynamical phases are identical for all $\left|w_{l,\mu}\right\rangle$ after $n_b$ cycles. Starting from general single-particle localized state $\left|\psi_j\right\rangle$ in flat band subspace, the evolved state is the initial state translated by $C$ unit cells,
\begin{equation}
	\begin{split}
		U(n_b T)\left|\psi_j\right\rangle=&\sum_{l,\mu}c_{l,\mu}\mathrm{e}^{-\mathrm{i}\theta_{l,\mu}(n_b T)}\left|{\cal W}_{l+C,\mu}\right\rangle, \\
		=&\mathrm{e}^{-\mathrm{i}\theta_{l,\mu}(n_b T)}\sum_{l,\mu}c_{l,\mu}\left|{\cal W}_{l+C,\mu}\right\rangle, \\
		=&\mathrm{e}^{-\mathrm{i}\theta_{l,\mu}(n_b T)} \mathcal{T}(Ca) \left|\psi_j\right\rangle.
	\end{split}
\end{equation}
And the off-diagonal term in Eq. \eqref{Eq:nb cycle displacement} vanishes, giving $\Delta x(n_b T)=Ca$. Consequently, the many-body state $\left|\Psi\right\rangle=\left|\psi_1,\psi_2,\dots,\psi_j,\dots\right\rangle$ exhibits charge pumping quantized to $C$ over $n_b$ cycles. However, this result is not robust against disorder, because dynamical phases do not differ by integer multiples of $2\pi$ when $E_{l^{\prime},\mu}-E_{l,\mu}$ deviates from $(l^{\prime}-l)Fa$.

\end{document}